\documentclass[twocolumn,showpacs,preprintnumbers,amsmath,amssymb,aps,floatfix]{revtex4}
\usepackage{txfonts}
\usepackage{graphics}
\usepackage{graphicx}% Include figure files
\usepackage{dcolumn}% Align table columns on decimal point
\usepackage{bm}% bold math
\usepackage{verbatim}
\usepackage{epsfig}
\usepackage{subfigure}
\usepackage{amsmath,amsfonts,amssymb,graphicx,times}
\usepackage{color}

\begin{document}

\title{Analytical and simulation studies of pedestrian flow at a crossing with random update rule}

\author{Zhong-Jun Ding$^{a,b}$}
\email{dingzj@hfut.edu.cn}
\author{Shao-Long Yu$^{a}$}
\author{Kongjin Zhu$^{a}$}
\author{Jian-Xun Ding$^{a}$}
\email{dingjianxun@hfut.edu.cn}
\author{Bokui Chen$^{c}$}
\author{Qin Shi$^{a}$}
\author{Rui Jiang$^{b}$}
\author{Bing-Hong Wang$^{d}$}

\affiliation{$^{a}$School of Automotive and Transportation Engineering, Hefei
University of Technology, Hefei 230009, People's Republic of China}

\affiliation{$^{b}$
 MOE Key Laboratory for Urban Transportation Complex Systems Theory and Technology,
Beijing Jiaotong University, Beijing 100044, People's Republic of China
}

\affiliation{$^{c}$
School of Computing, National University of Singapore, 117417, Singapore
}

\affiliation{$^{d}$
Department of Modern Physics, University of Science and Technology of China, Hefei 230026, People's Republic of China
}
\date{\today}

\begin{abstract}
The intersecting pedestrian flow on the 2D lattice with random update rule is studied. Each pedestrian has three moving directions without the back step. Under periodic boundary conditions, an intermediate phase has been found at which some pedestrians could move along the border of jamming stripes. We have performed mean field analysis for the moving and intermediate phase respectively. The analytical results agree with the simulation results well. The empty site moves along the interface of jamming stripes when the system only has one empty site. The average movement of empty site in one Monte Carlo step (MCS) has been analyzed through the master equation. Under open boundary conditions, the system exhibits moving and jamming phases. The critical injection probability $\alpha_{c}$ shows nontrivially against the forward moving probability $q$. The analytical results of average velocity, the density and the flow rate against the injection probability in the moving
phase also agree with simulation results well.


\end{abstract}


\maketitle

\section{Introduction}

Since serious trampling accidents always happen, pedestrian flow attracts more and more attention in recent years \cite{dyn1,dyn2}.
Understanding the properties of pedestrian flow is important for the design of urban facilities, traffic management and ensuring people's safety. Pedestrian dynamic has been studied in various fields including physics, engineering and mathematics.
Many basic and interesting phenomena  such as jamming, clogging, and lane formation have been observed \cite{dyn1,dyn2,con9,phen3}.

There are two main approaches for the study of pedestrian dynamics \cite{dyn1}. The first one is designing experiment or observing the real scenario through video \cite{exp2,exp3,exp4,exp5,exp6,exp7}. The other one is to describe the pedestrian flow by developing the delicate models.  These models include macroscopic and microscopic ones.
These macroscopic models are related to the traditional theory of fluid mechanics, etc \cite{Helbing1995,dyn1}.
Henderson had compared measurements of pedestrian flows with Navier-Stokes equations \cite{Henderson1}.

Microscopic models include social force, optimal velocity, cellular automata, lattice gas model, etc.
In some of these models such as the social force and optimal velocity models,  continuous time and space have been adopted \cite{con8,dyn1,exp5,con9,con10,nakayama}.
While some others are placed in a discretized time and space, such as cellular automata and lattice gas models \cite{cel11,cel12,cel13,cel14,cel15,cel16,pengy}.

The intersecting pedestrian flows are complex because of conflicts between two flows with different directions.
%Many studies are mainly carried out from experiments and models.
A number of field experiments on real intersecting pedestrian flows with four different angles have been conducted by Guo et al. \cite{exp17}. At the same time, a semi-continuous model has been developed and calibrated using
sample data. %in which pedestrian space is continuous and time is discrete.
Lian et al. \cite{exp18} have conducted a  series of controlled experiments of a four-directional intersecting pedestrian flow.
The average local velocity at high densities in the cross area is a bit larger than the previous study.
Muramatsu et al. have investigated the jamming transitions of pedestrian flow at a crossing under the periodic \cite{cel11} and open boundary conditions \cite{cel12} by the lattice gas model, respectively.
Hilhorst et al. have studied a lattice model of pedestrian traffic on two crossing
one-way streets \cite{mod19}. Its dynamics employs the frozen shuffle update.
Cividini et al. have explained stripe formation instability and revealed that the diagonal pattern actually consists of chevrons rather than straight diagonals \cite{mod27,mod28}.

The perpendicular traffic flow on two-dimensional lattice has been investigated by Biham et al. (BML ) using the cellular automaton model \cite{mod21}.  Except the moving and jamming phase, D'Souza have found an intermediate stable phase with free-flowing regions intersecting at
jammed wave fronts in the original BML model \cite{inter phase}.
Ding et al. have studied an stochastic BML model with random update rule (BML-R) \cite{mod22}.
A phase separation phenomenon has been observed when the slow-to-start effect in the BML model is considered \cite{suiqh}.

Almost all of the models presented above are random-sequential, sublattice-parallel or parallel while the model with random update procedures is scarce. Since the pedestrians always behave randomly in real life, the model with random update procedures is considered in this paper. This paper investigate effects of the random update procedures on the properties of the stationary state of intersecting pedestrian flows.
An intermediate phase where some pedestrians move along the border of jamming stripes was found.
The average velocity of the moving and intermediate phase have been analyzed through the mean filed analysis. The analytical results are in good agreement with the simulation ones. The empty site moves along the interface of jamming stripes when the system only has one empty site. The average movement of the empty site in one MCS was analyzed through the master equation.  The critical injection probability $\alpha_{c}$ under open boundary conditions shows nontrivially against the forward moving probability $q$.

This paper is organized as follows: The models are introduced in Section 2. In Section 3, we compare the analytical results with simulation ones under periodic and open boundary conditions, in 3.1 and 3.2, respectively. Section 4 gives the conclusions.

\section{Model}

\begin{figure}
    \begin{center}
    \scalebox{0.3}[0.3]{\includegraphics{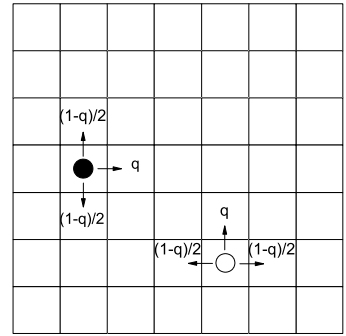}}
    \caption{Illustration of the intersecting pedestrian flow at a crossing. The E and N pedestrians are represented by the solid and open circles, respectively .
}\label{Figure 1}
    \end{center}
\end{figure}

There are two species of pedestrians distributed randomly on a 2D square lattice $L\times L$ with the same densities. As shown in FIG. \ref{Figure 1} each pedestrian moves to the preferential direction with no backstep. The first (second) type of pedestrian is eastbound (northbound), E pedestrian (N pedestrian) for short. For example, the E pedestrian could move to the eastward, northward and southward site while the N pedestrian could move to the northward, eastward and westward site.
The pedestrians exclude each other on a site. Thus, each lattice site can be in one of three states: empty, occupied by the E pedestrian, or occupied by the N pedestrian.
%For example, the first type of pedestrians select eastward, northward and southward site with probability $q$, $(1-q)/2$ and $(1-q)/2$. The second type select northward, eastward and westward site with probability$q$, $(1-q)/2$ and $(1-q)/2$.

Under periodic boundary conditions, the following steps are repeated $L^2$ times in one Monte Carlo step (MCS): (i) one site is selected randomly; (ii) if the selected site is empty, nothing happens; otherwise, if the selected site is occupied by the E pedestrian, the nearest east, south and north neighboring site are chosen as target site with probabilities  $q$, $(1-q)/2$ and $(1-q)/2$, respectively;
% the pedestrian has three moving directions: to the nearest east, the nearest north and the nearest south. Three probabilities of the pedestrian moving to the neighbor sites are $q$, $(1-q)/2$ and $(1-q)/2$ unless the target site is occupied.
 otherwise, if the selected site is occupied by the N pedestrian, the nearest north, east and west neighboring site are chosen as target site with probabilities $q$, $(1-q)/2$ and $(1-q)/2$;
(iii) the pedestrian moves to the target site unless it is occupied.

 %the pedestrian has three moving directions: to the nearest north, to the nearest west and to the nearest east. Three probabilities of the pedestrian moving to the neighbor sites are $q$, $(1-q)/2$ and $(1-q)/2$ unless the target site is occupied.

Under open boundary conditions, the E (N) pedestrians are injected with probability $\alpha$ on the west (south) boundary and removed with probability $\beta$ on the other three boundaries. At the southwest corner, the E or N pedestrians are injected with probability $\alpha/2$.

%Fig.1

\section{Results}

\subsection{Periodic boundary}
\subsubsection{Simulation results}

The lattice size is set as $100\times100$ unless otherwise mentioned. For each density, we simulate 100 runs. The result of each run is obtained after discarding the first $10^6$ MCSs (as transient time) and averaged in the next $10^5$ MCSs.

%Fig.2
\begin{figure}
    \begin{center}
    \scalebox{0.3}[0.3]{\includegraphics{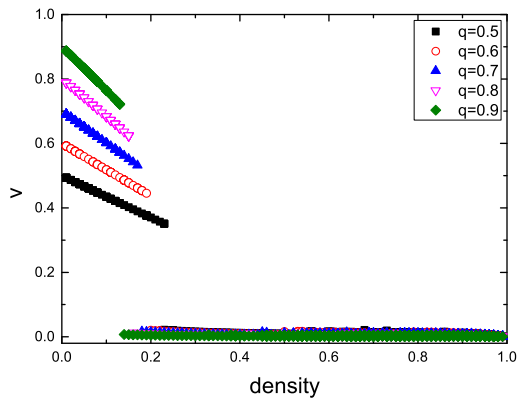}}
    \caption{The average velocity of each run against the pedestrian density $\rho$ as $q= 0.5, 0.6, 0.7, 0.8, 0.9$ for the lattice size $100 \times 100$.}\label{Figure 2}
    \end{center}
\end{figure}

The average velocity $v$ of each run against the pedestrian density $\rho$ for $q = 0.5, 0.6, 0.7, 0.8, 0.9$ are shown in FIG. \ref{Figure 2}. The average velocity $v$ is defined as the average directed distance divided by the MCS. When $q=1$, our model reduces to the BML-R model \cite{mod22}. One can see that two phases could be observed, i.e., the moving and intermediate phase. In the moving phase, all pedestrians can move. In the intermediate phase, the average velocity becomes a non-negligible small value $v>0$ instead of $v=0$. There is a range of densities in which the two phases coexist and we denote the center of this range as $\rho_{c}$. One can see that with the increase of $q$, the critical density $\rho_{c}$ decreases.

\begin{figure}
    \begin{center}
    \scalebox{0.3}[0.3]{\includegraphics{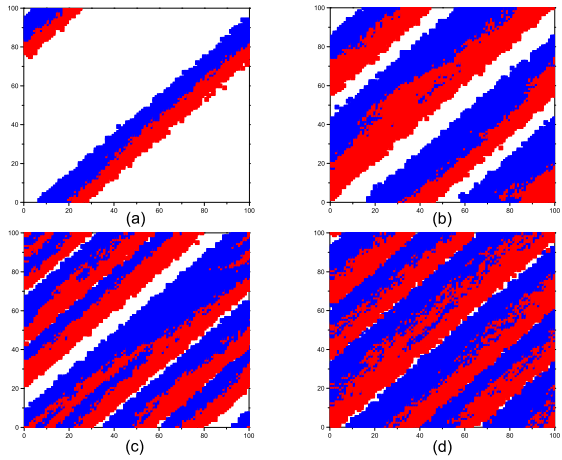}}
    \caption{Three typical configurations of the intermediate phase. The parameters are $L$ = 100,  and (a)$q$=0.6, $\rho$=0.2,(b)$q$=0.6, $\rho$=0.7, (c)$q$=0.8, $\rho$=0.7,(d)$q$=0.6, $\rho$=0.9.  The E pedestrian is indicated by blue and the N pedestrian is indicated by red.}\label{Figure 3}
    \end{center}
\end{figure}

The three typical configurations for the intermediate phase are shown in FIG. \ref{Figure 3}. Some pedestrians are stopped at the interior of the  cluster (stripe) while others could move along the border of the jamming stripes from the lower left to the upper right corner. With the increase of density and $q$, the number of stripes in one row (column) increases (see FIG. \ref{Figure 3}(b) and (c)).

\subsubsection{Analytical result of moving phase}

 We have developed a mean field analysis for the average velocity in the moving phase by extending the method of reference \cite{mod22}. The E (N) pedestrian could move to the east, the north or the south (the north, the east or the west) site with different probabilities. If we have selected a site occupied by the E (N) type of pedestrian, then the probability that its east (north) site is empty is assumed to be $p_{f}$, while the probability that its north or south (west or east) site is empty is assumed to be $p_{s}$.

\begin{figure}
    \begin{center}
    %\scalebox{0.3}[0.3]{\includegraphics{4-1.eps}}
    \end{center}
\end{figure}

\begin{figure}*
    \begin{center}
    \scalebox{0.3}[0.3]{\includegraphics{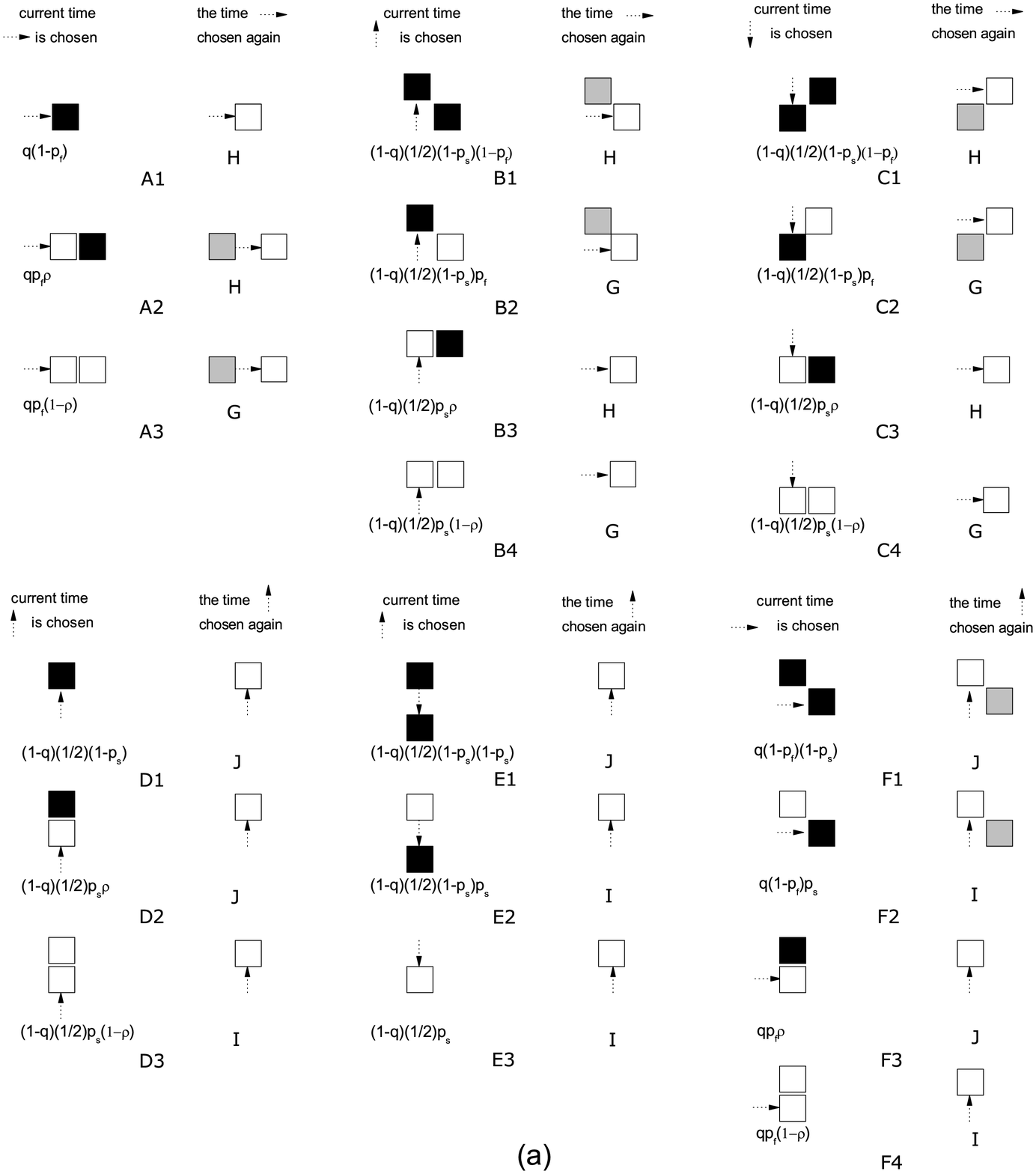}}
    \scalebox{0.3}[0.3]{\includegraphics{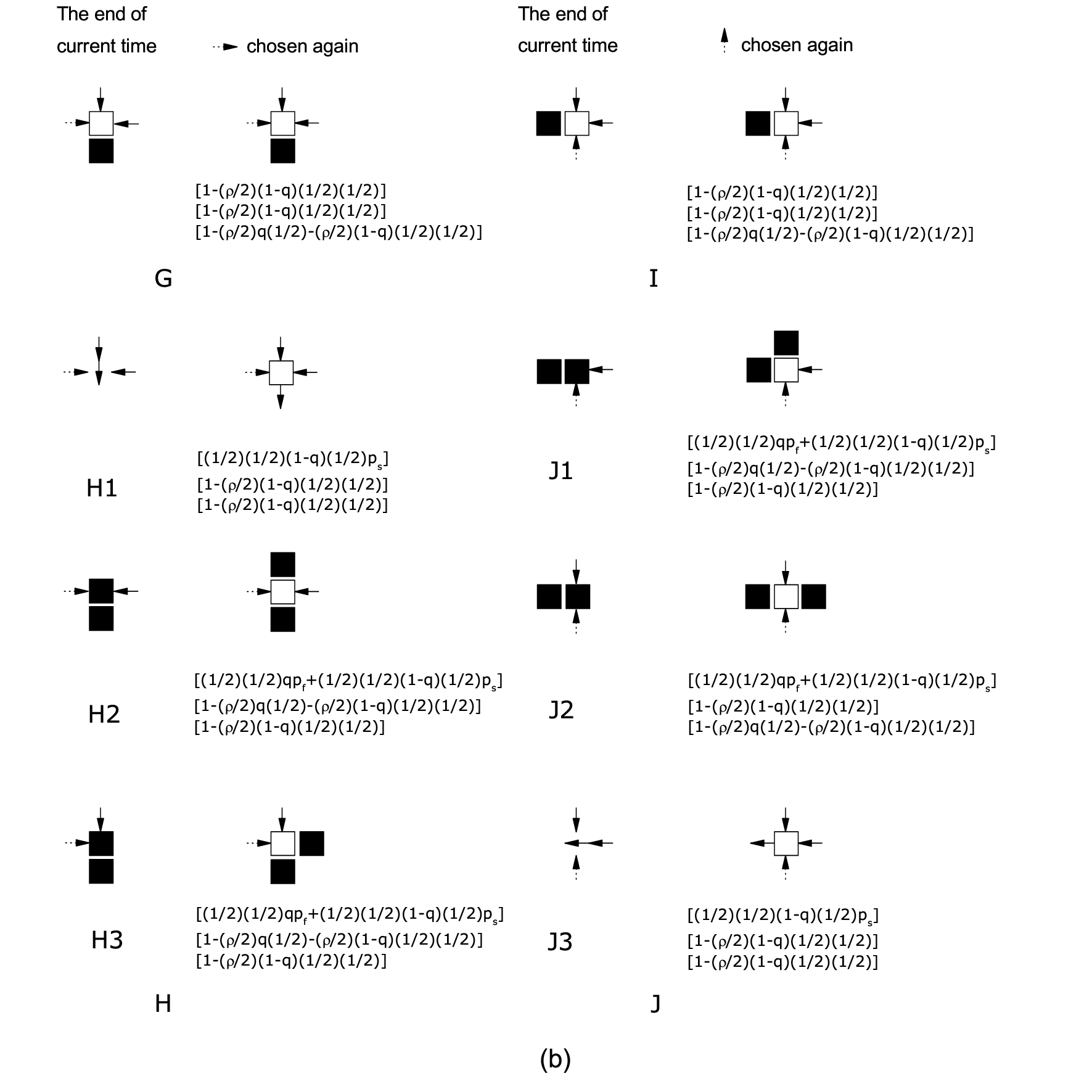}}
    \caption{The illustration of the mean field method. The dashed arrows represents the E pedestrian. The solid arrows $\downarrow$ ($\leftarrow$) only represents the E (N) pedestrian moving south (west). The dark, empty and gray box represent a pedestrian, an empty site and a site which is either empty or occupied, respectively.}\label{Figure 4}
    \end{center}
\end{figure}

 In order to calculate these $p_{f}$ and $p_{s}$, 21 situations have been considered as shown in FIG. \ref{Figure 4}$(a)$. The dashed arrows represents the target pedestrian. Since the symmetry of the model, the target pedestrian is restricted to the E pedestrian. The solid arrows $\downarrow$ ( $\leftarrow$ ) only represents the E (N) pedestrian moving south (west). The dark and empty box represent a pedestrian and an empty site, respectively.

The left side of each subfigure is the existence probability of the corresponding situation at the current time. The right side is the situation that the target pedestrian could move when is chosen again. The moving probability of the target pedestrian when is chosen again is shown in the subfigures $G$, $H$, $I$ and $J$ respectively.

We explain the subfigures $A1$, $A3$, $B1$, $D1$, $H$ and $G$ in detail. Other situations could be obtained similarly.

\textbf{Subfigure $A1$:}

On the left side of subfigure $A1$, $q$ is the probability that the target pedestrian moves east and $(1-p_{f})$ is the probability that the east site of $\dashrightarrow$ is occupied so that $\dashrightarrow$ could not move in the current time step. If the target pedestrian wants to move to the east site successfully when it is chosen again, its east site must be empty. The probability that the east site of $\dashrightarrow$ becomes empty is shown in the subfigure $H$.

\textbf{Subfigure $H$:}

The left side of subfigure $H$ is the configuration at the end of the current time of subfigure $A1$. On the right site of $H1$, the three terms in $[\cdots]$ corresponds to that of east, north-east and east-east site, respectively.

The first (1/2) in the first $[\cdots]$ is the probability that the east site of $\dashrightarrow$ is occupied by the E pedestrian, the second (1/2) is the probability that the E pedestrian is chosen before $\dashrightarrow$ is chosen again, $(1-q)(1/2)$ is the probability that the E pedestrian moves south and $p_{s}$ is the probability that the south-east site of $\dashrightarrow$ is empty. So that $[(1/2)(1/2)(1-q)(1/2)p_{s}]$ is the probability that the east site of $\dashrightarrow$ becomes empty first.

The $(\rho/2)$ in the second $[\cdots]$ is the probability that the northeast site of $\dashrightarrow$ is occupied by the E pedestrian, $(1-q)(1/2)(1/2)$ is the probability that the E pedestrian in the northeast site is chosen and moves south before $\dashrightarrow$ is chosen again. Similarly, $(\rho/2)(1-q)(1/2)(1/2)$ in the third $[\cdots]$ represents that the east-east site is occupied by the N pedestrian and it moves west before $\dashrightarrow$ is chosen again.

Thus $[1-(\rho/2)(1-q)(1/2)(1/2)][1-(\rho/2)(1-q)(1/2)(1/2)]$ is the probability that none of the pedestrian at the northeast or east-east site moves to the empty east site of $\dashrightarrow$. So the east site of $\dashrightarrow$ remains empty after the pedestrian in the east site moves out.

The probabilities in subfigures $H2$ and $H3$ could be obtained similarly. Since the east site could be occupied by either the first or the N pedestrian in subfigure $H2$. Thus  $(1/2)(1/2)qp_{f}$ and $(1/2)(1/2)(1-q)(1/2)p_{s}$ in the first $[\cdots]$ corresponds to that of the N pedestrian and E pedestrian, respectively.

\textbf{Subfigure $A3$:}

Similarly, on the left of subfigure $A3$, $q$ is the same as that of $A1$, $p_{f}$ is the probability that the east site of $\dashrightarrow$ is empty so that $\dashrightarrow$ could move in the current time step, $(1-\rho)$ is the probability that the east-east site of $\dashrightarrow$ is empty. If the target pedestrian wants to move to its east-east site successfully when it is chosen again, the east-east site must remain empty. The probability that the east-east site stays empty is shown in the subfigure $G$.

\textbf{Subfigure $G$:}

The left side of subfigure $G$ is the configuration at the end of current time of subfigures $A3$. On the right side of subfigure $G$, the three terms in $[\cdots]$ corresponds to that of east-east,  north-east sites and south-east site, respectively.

The $(\rho/2)$ in the first $[\cdots]$ is the probability that the east-east site of $\dashrightarrow$ is occupied by the N pedestrian. $(1-q)(1/2)(1/2)$ is the probability that the N pedestrian in the east-east site is chosen and moves west before $\dashrightarrow$ is chosen again. Similarly, the second $[\cdots]$ corresponds to that of north-east sites.

Since the south-east site could be occupied by either the E pedestrian or the N pedestrian. Thus $(\rho/2)q(1/2)$ and $(\rho/2)(1-q)(1/2)(1/2)$ in the third $[\cdots]$ corresponds to the N pedestrian and the E pedestrian, respectively.

Thus [1-($\rho$/2)$(1-q)$(1/2)(1/2)][1-($\rho$/2)$(1-q)$(1/2)(1/2)][1-($\rho$/2)$q$(1/2)-($\rho$/2)$(1-q)$(1/2)(1/2)] is the probability that none of pedestrians moves to the east site of $\dashrightarrow$ before $\dashrightarrow$ is chosen again.

\textbf{Subfigure $B1$:}

On the left side of subfigure $B1$, $(1-q)(1/2)$ is the probability that the E pedestrian moves north. $(1-p_{s})$ and $(1-p_{f})$ are the probabilities that the north and east site of $\uparrow$ are occupied, respectively. So that $\uparrow$ could not move in the current time step. If the target pedestrian wants to move to the east site successfully when it is chosen again, the east site must be empty. The probability that the east site becomes empty is also shown in the subfigure $H$.

\textbf{Other subfigures:}

The subfigures $D1\sim D3$, $E1\sim E3$ and $F1\sim F4$ are similar to the subfigures $A1\sim A3$, $B1\sim B4$ and $C1\sim C4$ respectively. On the left side of subfigure $D1$, $(1-q)(1/2)$ is the probability that the E pedestrian moves north and $(1-p_{s})$ is the probability that the north site of $\uparrow$ is occupied so that $\uparrow$ could not move north in the current time step. If the target pedestrian wants to move to the north site successfully when it is chosen again, the north site must be empty. The probability that the north site becomes empty is shown in the subfigure $J$. The meanings of the terms in $J$ are similar to that of $H$.

Since the E pedestrian can move east when it is chosen again in the situation $A1\sim A3$, $B1\sim B4$ and $C1\sim C4$ and north in $D1\sim D3$, $E1\sim E3$ and $F1\sim F4$ , thus we have following two equations :

\begin{widetext}
\begin{eqnarray}\label{1}
p_{f}&=&[q(1-p_{f})+qp_{f}\rho+\frac{1}{2}(1-q)(1-p_{s})(1-p_{f})+\frac{1}{2}(1-q)p_{s}\rho+\frac{1}{2}(1-q)(1-p_{s})(1-p_{f})
{}\nonumber\\& &
+\frac{1}{2}(1-q)p_{s}\rho]\lbrace[\frac{1}{2}\frac{1}{2}(1-q)\frac{1}{2}p_{s}][1-\frac{\rho}{2}(1-q)\frac{1}{2}\frac{1}{2}]
[1-\frac{\rho}{2}(1-q)\frac{1}{2}\frac{1}{2}]+[\frac{1}{2}\frac{1}{2}qp_{f}
{}\nonumber\\& &
+\frac{1}{2}\frac{1}{2}(1-q)\frac{1}{2}p_{s}][1
-\frac{\rho}{2}q\frac{1}{2}-\frac{\rho}{2}(1-q)\frac{1}{2}\frac{1}{2}][1-\frac{\rho}{2}(1-q)\frac{1}{2}\frac{1}{2}]
+[\frac{1}{2}\frac{1}{2}qp_{f}+\frac{1}{2}\frac{1}{2}(1-q)\frac{1}{2}p_{s}][1
{}\nonumber\\& &
-\frac{\rho}{2}q\frac{1}{2}-\frac{\rho}{2}(1-q)\frac{1}{2}\frac{1}{2}][1-\frac{\rho}{2}(1-q)\frac{1}{2}\frac{1}{2}]\rbrace
+[qp_{f}(1-\rho)+\frac{1}{2}(1-q)(1-p_{s})p_{f}+\frac{1}{2}(1-q)p_{s}(1-\rho)
{}\nonumber\\& &
+\frac{1}{2}(1-q)(1-p_{s})p_{f}+\frac{1}{2}(1-q)p_{s}(1-\rho)]\lbrace[1-\frac{\rho}{2}(1-q)\frac{1}{2}\frac{1}{2}]
[1-\frac{\rho}{2}(1-q)\frac{1}{2}\frac{1}{2}][1-\frac{\rho}{2}q\frac{1}{2}-\frac{\rho}{2}(1-q)\frac{1}{2}\frac{1}{2}]\rbrace.
\end{eqnarray}

\begin{eqnarray}\label{2}
p_{s}&=&[\frac{1}{2}(1-q)(1-p_{s})+\frac{1}{2}(1-q)p_{s}\rho+\frac{1}{2}(1-q)(1-p_{s})(1-p_{s})
+q(1-p_{f})(1-p_{s})+qp_{f}\rho]\lbrace[\frac{1}{2}\frac{1}{2}qp_{f}
{}\nonumber\\& &
+\frac{1}{2}\frac{1}{2}(1-q)\frac{1}{2}p_{s}][1-\frac{\rho}{2}q\frac{1}{2}-\frac{\rho}{2}(1-q)\frac{1}{2}\frac{1}{2}][1
-\frac{\rho}{2}(1-q)\frac{1}{2}\frac{1}{2}]+[\frac{1}{2}\frac{1}{2}qp_{f}+\frac{1}{2}\frac{1}{2}(1-q)\frac{1}{2}p_{s}][1
{}\nonumber\\& &
-\frac{\rho}{2}(1-q)\frac{1}{2}\frac{1}{2}][1-\frac{\rho}{2}q\frac{1}{2}-\frac{\rho}{2}(1-q)\frac{1}{2}\frac{1}{2}]
+[\frac{1}{2}\frac{1}{2}(1-q)\frac{1}{2}p_{s}][1-\frac{\rho}{2}(1-q)\frac{1}{2}\frac{1}{2}][1-\frac{\rho}{2}(1-q)\frac{1}{2}\frac{1}{2}]\rbrace
{}\nonumber\\& &
+[\frac{1}{2}(1-q)p_{s}(1-\rho)+\frac{1}{2}(1-q)(1-p_{s})p_{s}+\frac{1}{2}(1-q)p_{s}+q(1-p_{f})p_{s}
{}\nonumber\\& &
+qp_{f}(1-\rho)]\lbrace[1-\frac{\rho}{2}(1-q)\frac{1}{2}\frac{1}{2}][1-\frac{\rho}{2}(1-q)\frac{1}{2}\frac{1}{2}]
[1-\frac{\rho}{2}q\frac{1}{2}-\frac{\rho}{2}(1-q)\frac{1}{2}\frac{1}{2}]\rbrace.
\end{eqnarray}
\end{widetext}

The average velocity can be obtained from $p_{f}$,
\begin{eqnarray}\label{3}
<v>&=&qp_{f}.
\end{eqnarray}

\begin{figure}
    \begin{center}
    \scalebox{0.3}[0.3]{\includegraphics{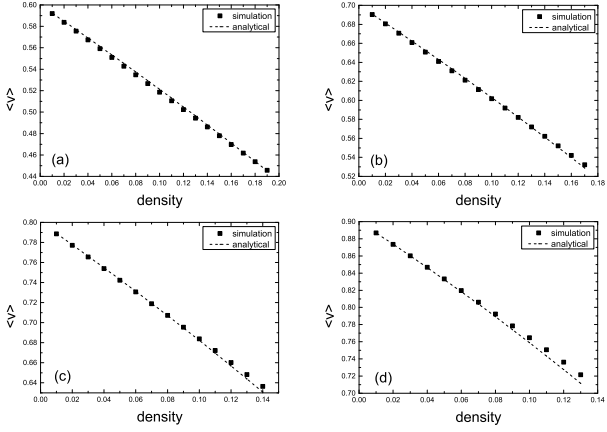}}
    \caption{The comparison of the simulation results with the mean field analysis for the moving phase. The dashed lines are analytical results and black squares are simulation ones. The average velocity $<v>$ are the average of 100 runs for each density. (a)$q$=0.6, (b)$q$=0.7, (c)$q$=0.8, (d)$q$=0.9.}\label{Figure 5}
    \end{center}
\end{figure}

FIG. \ref{Figure 5} compares the simulation results with the analytical results. One can see that the analytical results are in good agreement with the simulation results, which proves that our analytical method is effective.

\subsubsection{Analytical results of intermediate phase}

\begin{figure}
    \begin{center}
    \scalebox{0.3}[0.3]{\includegraphics{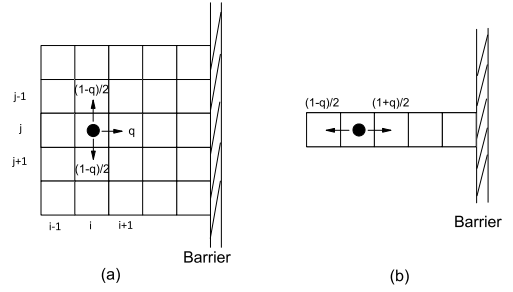}}
    \caption{The 2D model in the intermediate phase is transformed into the 1D model, which is similar to that of reference \cite{mod24,mod25,mod26}.}\label{Figure 6}
    \end{center}
\end{figure}

We have developed two methods to analyse the average velocity $v$ in the intermediate phase. As shown in FIG.  \ref{Figure 6}, the solid circle represents the E pedestrian, whose coordinates are $(i,j)$. Then its east, north and south neighbors' coordinates are $(i+1,j)$, $(i,j-1)$ and $(i,j+1)$,respectively.

The master equation (4) is used in the first method

\begin{eqnarray}\label{4}
\frac{d<\tau_{(i,j)}>}{dt}&=&\lbrace q<\tau_{(i-1,j)}(1-\tau_{(i,j)})>
{}\nonumber\\& &
+(1-q)(1/2)<\tau_{(i,j-1)}(1-\tau_{(i,j)})>
{}\nonumber\\& &+(1-q)(1/2)<\tau_{(i,j+1)}(1-\tau_{(i,j)})>\rbrace
{}\nonumber\\& &
-\lbrace q<\tau_{(i,j)}(1-\tau_{(i+1,j)})>
{}\nonumber\\& &
+(1-q)(1/2)<\tau_{(i,j)}(1-\tau_{(i,j-1)})>
{}\nonumber\\& &
+(1-q)(1/2)<\tau_{(i,j)}(1-\tau_{(i,j+1)})>\rbrace
{}\nonumber\\
&=&0,
\end{eqnarray}
where $\tau_{(i,j)}=$ 0 or 1 is the occupation number at site $(i,j)$. $\tau_{(i,j)}=1$ means that site $(i,j)$ is occupied while $\tau_{(i,j)}=0$ means empty. $<\cdots>$ represents an average with respect to all the microstates. The terms $q<\tau_{(i-1,j)}(1-\tau_{(i,j)})>$, $(1-q)(1/2)<\tau_{(i,j-1)}(1-\tau_{(i,j)})>$ and $(1-q)(1/2)<\tau_{(i,j+1)}(1-\tau_{(i,j)})>$ represent a flux into site $(i,j)$. While $q<\tau_{(i,j)}(1-\tau_{(i+1,j)})>$, $(1-q)(1/2)<\tau_{(i,j)}(1-\tau_{(i,j-1)})>$ and $(1-q)(1/2)<\tau_{(i,j)}(1-\tau_{(i,j+1)})>$ represent a flux out of site $(i,j)$.

Since the jamming stripes are oriented along the diagonal direction, $<\tau_{(i,j-1)}> \approx <\tau_{(i-1,j)}>$ and $<\tau_{(i,j+1)}> \approx <\tau_{(i+1,j)}>$ are assumed. Equation (4) becomes

\begin{eqnarray}\label{5}
\frac{d<\tau_{(i,j)}>}{dt}&=&(1-q)(1/2)\lbrace <\tau_{(i+1,j)}>(1-<\tau_{(i,j)}>)
{}\nonumber\\& &
-<\tau_{(i,j)}>(1-<\tau_{(i-1,j)}>)\rbrace
{}\nonumber\\& &
+(1+q)(1/2)\lbrace <\tau_{(i-1,j)}>(1-<\tau_{(i,j)}>)
{}\nonumber\\& &
-<\tau_{(i,j)}>(1-<\tau_{(i+1,j)}>)\rbrace
{}\nonumber\\
&=& 0.
\end{eqnarray}
where the approximation $<\tau_{i}\tau_{i+1}> = <\tau_{i}><\tau_{i+1}>$ is used.
From equation (5) one can see that flux only occurs among sites $(i,j)$, $(i-1,j)$ and $(i+1,j)$, ie., the horizontal direction. The two dimensional model (FIG. \ref{Figure 6}(a)) is reduced to one dimension model (FIG. \ref{Figure 6}(b)). The density distribution of intermediate phase in one row (column) is similar to that of reference\cite{mod24,mod25,mod26}.

Therefore equation (6) can be obtained by solving equation (5),

\begin{eqnarray}\label{6}
<\tau_{i+1}>&=&\frac{2<\tau_{i}>+2q<\tau_{i}><\tau_{i-1}>}{2<\tau_{i}>+1-q},
\end{eqnarray}
in which the label $j$ is taken off. The density of each site $<\tau_{i}>$ could be obtained by the iteration of equation (6). The densities of the first and second site adopt $<\tau_{0}> = 0$ and $<\tau_{1}> \to 0$ respectively for the beginning of iteration. The average eastward movement for the 1D lattice in one MCS $v_{1}$ could be calculated by the following equation

\begin{eqnarray}\label{7}
v_{1}&=&\sum_{i=1}^{\infty}q<\tau_{i}>(1-<\tau_{i+1}>).
\end{eqnarray}

There are $n_{s}\times L$ 1D lattice for the E pedestrian and the N pedestrian respectively when the system has $n_{s}$ stripes. The average velocity of the intermediate phase is

\begin{eqnarray}\label{8}
v&=&\frac{n_{s}Lv_{1}}{(\rho/2)L^2}.
\end{eqnarray}

The second method uses the results of reference \cite{mod24,mod25} for the 1D lattice in FIG. \ref{Figure 6}(b),

\begin{eqnarray}\label{9}
p(\tau_{1},\cdots,\tau_{i},\cdots,\tau_{L})&=&\frac{1}{Z}\prod_{i}(\frac{q'}{1-q'})^{i\tau_{i}}
{}\nonumber\\
&=&\frac{1}{Z}(\frac{q'}{1-q'})^{\sum_{i=1}^{i=L}i\tau_{i}}
{}\nonumber\\
&=&\frac{1}{Z}e^{ln(\frac{q'}{1-q'})^{\sum_{i=1}^{i=L}i\tau_{i}}},
\end{eqnarray}
where $p(\tau_{1},\cdots,\tau_{i},\cdots,\tau_{L})$ denotes the probability density of microstates $(\tau_{1},\cdots,\tau_{i},\cdots,\tau_{L})$ and $Z$ is a normalization factor, such that the sum of $p(\tau_{1},\cdots,\tau_{i},\cdots,\tau_{L})$ over all allowed configurations is 1. Here , $q'$ is the probability the pedestrian moving to the right while $1-q'$ to the left in the 1D lattice. From FIG. \ref{Figure 6}(b), one can see that $q'=(1+q)/2$ and $1-q'=(1-q)/2$ in our model. The equation (9) becomes

\begin{eqnarray}\label{10}
p(\tau_{1},\cdots,\tau_{i},\cdots,\tau_{L})&=&\frac{1}{Z}e^{ln(\frac{1+q}{1-q})^{\sum_{i=1}^{i=L}i\tau_{i}}}.
\end{eqnarray}
Since the exclusion property of particles, the density of each site satisfies the Fermi distribution

\begin{eqnarray}\label{11}
<\tau_{i}>&=&\frac{1}{e^{-(i-u)ln(\frac{1+q}{1-q})}+1},
\end{eqnarray}
where $\mu \gg 0$. When $(i-\mu)$ in equation (11) is substituted by $i$, i.e. $<\tau_{i}>=\frac{1}{e^{-(i)ln(\frac{1+q}{1-q})}+1}$,then equation (7) becomes

\begin{eqnarray}\label{12}
v_{1}&=&\sum_{i=-\infty}^{\infty}q<\tau_{i}>(1-<\tau_{i+1}>).
\end{eqnarray}

Since

\begin{eqnarray}\label{13}
\sum_{i=-\infty}^{\infty}f(i)&=&\sum_{i=-\infty}^{\infty}\int_{-\infty}^{\infty}f(x)\delta(x-i)dx
{}\nonumber\\
&=&\int_{-\infty}^{\infty}f(x)\sum_{i=-\infty}^{\infty}\delta(x-i)dx
{}\nonumber\\
&=&\int_{-\infty}^{\infty}f(x)\sum_{m=-\infty}^{\infty}e^{2\pi imx}dx
{}\nonumber\\
&=&\sum_{m=-\infty}^{\infty}\int_{-\infty}^{\infty}f(x)e^{2\pi imx}dx,
\end{eqnarray}

then

\begin{eqnarray}\label{14}
v_{1}&=&\sum_{m=-\infty}^{\infty}\int_{-\infty}^{\infty}e^{2\pi imx}q<\tau(x)>(1-<\tau(x+1)>)dx.
{}\nonumber\\& &
\end{eqnarray}

Compared to the term $m = 0$ in equation (14), other terms $m\neq 0$ could be neglected. Thus

\begin{eqnarray}\label{15}
v_{1}&\approx&
\int_{-\infty}^{\infty}q<\tau(x)>(1-<\tau(x+1)>)dx
{}\nonumber\\
&\approx&\frac{1-q}{2}.
\end{eqnarray}

Substituting equation (15) into equation(16), the average velocity of the intermediate phase is

\begin{eqnarray}\label{16}
v&=&
\frac{n_{s}Lv_{1}}{(\rho/2)L^2}
=\frac{n_{s}(1-q)}{\rho L}.
\end{eqnarray}

\begin{figure}
    \begin{center}
    \scalebox{0.3}[0.3]{\includegraphics{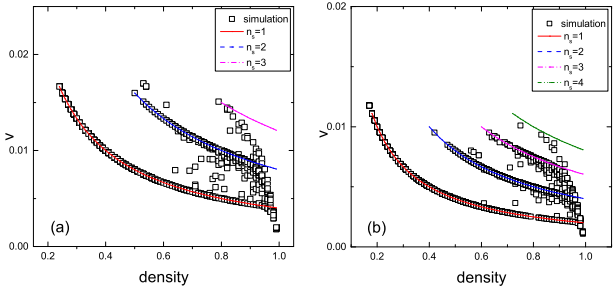}}
    \caption{The comparison of the simulation results with the mean field analysis for the intermediate phase as $q$=(a) 0.6, (b) 0.8 for the lattice size $100 \times 100$. $v$ is the average velocity of each run.}\label{Figure 7}
    \end{center}
\end{figure}

FIG. \ref{Figure 7} compares the simulation results with the mean field results. One can see that the analytical results are in good agreement with the simulation results. As shown in FIG. \ref{Figure 3}(d), when the density is in the high density region of the intermediate phase, the jamming stripes meet and merge with each other, which makes equation (16) not suitable. One can see that with the increasing of $q$, the density where multi stripes begin to appear decreases.

\subsubsection{The analytic result when the system only has one empty site}
\begin{figure}
    \begin{center}
    \scalebox{0.3}[0.3]{\includegraphics{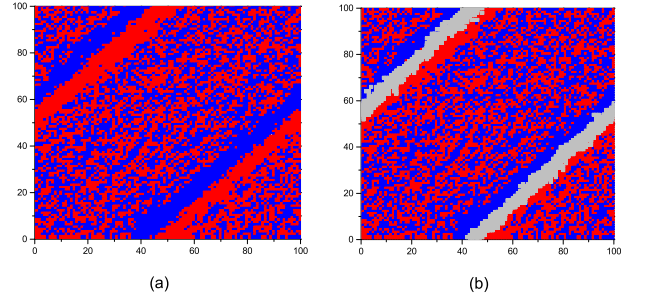}}
    \caption{(a) The typical configuration when the system only has one empty site as $q=0.6$ ; (b) The moving trajectory of empty site in (a). The blue (red) represents the E (N) pedestrian, and the gray indicates the trajectory of the empty site.}\label{Figure 8}
    \end{center}
\end{figure}

When the system has only one empty site, the model can be seen as the random walk of the empty site on the 2D lattice. However, a self-organized pattern with stripes along the diagonals from the upper-right to the lower-left corners was observed (see FIG. \ref{Figure 8}(a)). FIG. \ref{Figure 8}(b) shows the moving trajectory of the empty site in $10,000$ MCS. One can see that the empty site moves along the interface of jamming stripe. Next, we analyse the average movement $<v_{mov}>$ of empty site in one MCS. As shown in FIG 9(a), according to the neighbors' states of empty site, there are 16 possible configurations, i.e. $S_{1}^{a}\sim S_{16}^{a}$. The probabilities of configurations $S_{1}^{a}\sim S_{16}^{a}$ are denoted as $p_{1}\sim p_{16}$ , respectively.

We develop an analytical method to solve $p_{1}\sim p_{16}$. The evolution process of $S_{1}^{a}$ is shown in FIG. \ref{Figure 10}(a). One can see that $S_{1}^{a}$ can be evolved from three configurations $S_{1}^{c}$, $S_{2}^{c}$ and $S_{3}^{c}$.  The corresponding transition probabilities are $q$, $(1-q)/2$ and $(1-q)/2$, respectively.  While $S_{1}^{a}$ can also evolve into other three configurations $S_{1}^{d}$, $S_{2}^{d}$ and $S_{3}^{d}$. The corresponding transition probabilities are $q$, $(1-q)/2$ and $(1-q)/2$, respectively. Thus, the master equation for configurations $S_{1}^{a}$ is

\begin{figure}
    \begin{center}
    \scalebox{0.3}[0.3]{\includegraphics{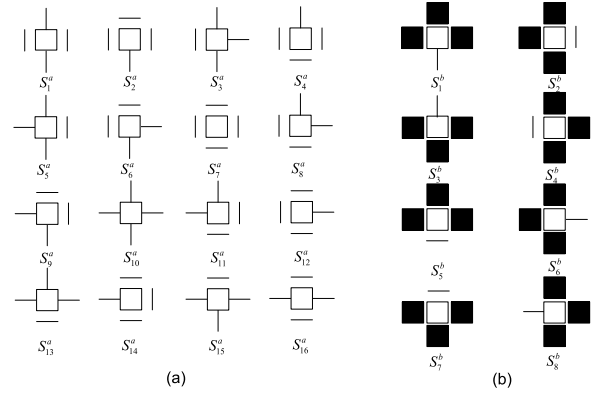}}
    \caption{(a) Sixteen typical configurations of the empty site's neighbors. $-$ ($\lvert$) represents the E (N) pedestrian. (b) Eight typical configurations of the empty site's neighbors. The solid box represents a pedestrian which is either the E pedestrian or the N pedestrian.}\label{Figure 9}
    \end{center}
\end{figure}

\begin{figure}
    \begin{center}
    \scalebox{0.3}[0.3]{\includegraphics{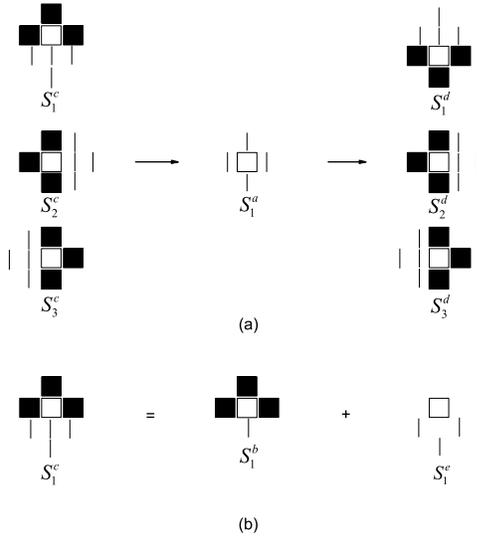}}
    \caption{(a) The evolution process of $S_{1}^{a}$. (b) The configurations $S_{1}^{c}$ is composed of $S_{2}^{b}$ and $S_{1}^{e}$. $-$ ($\lvert$) represents the E (N) pedestrian.}\label{Figure 10}
    \end{center}
\end{figure}

\begin{eqnarray}\label{17}
\frac{dp_{1}}{dt}&=&[p_{S_{1}^{c}}q+p_{S_{2}^{c}}(1-q)(1/2)+p_{S_{3}^{c}}(1-q)(1/2)]
{}\nonumber\\& &
-[p_{1}q+p_{1}(1-q)(1/2)+p_{1}(1-q)(1/2)]
{}\nonumber\\&=&0,
\end{eqnarray}
where $p_{S_{1}^{c}}$ , $p_{S_{2}^{c}}$ and $p_{S_{3}^{c}}$ are the probability of configurations $S_{1}^{c}$, $S_{2}^{c}$ and $S_{3}^{c}$, respectively.

From FIG. \ref{Figure 10}(b) one can see that the configurations $S_{1}^{c}$ is composed of $S_{1}^{b}$ and $S_{1}^{e}$. We assume that

\begin{eqnarray}\label{18}
p_{S_{1}^{c}}&=&p_{S_{1}^{b}}p(S_{1}^{c}|S_{1}^{b})\approx p_{S_{1}^{b}}p_{S_{1}^{e}},
\end{eqnarray}
where $p_{S_{1}^{b}}$ and $p_{S_{1}^{e}}$ are the probabilities of configurations $S_{1}^{b}$ and $S_{1}^{e}$, respectively.

As shown in FIG. \ref{Figure 9}(b), $S_{1}^{b}$ represents that the south neighbor of the empty site is the N pedestrian and the other three neighbors can be the E pedestrian or the N pedestrian. Therefore, the probability of $S_{1}^{b}$ is the sum of the probabilities of the eight typical configurations $S_{1}^{a}, S_{2}^{a}, S_{3}^{a}, S_{5}^{a}, S_{6}^{a}, S_{9}^{a}, S_{10}^{a}$ and $S_{15}^{a}$, i.e.,

\begin{eqnarray}\label{19}
p_{S_{1}^{b}}&=&p_{1}+p_{2}+p_{3}+p_{5}+p_{6}+p_{9}+p_{10}+p_{15}.
\end{eqnarray}
These $p_{S_{2}^{b}}\sim p_{S_{8}^{b}}$ can be obtained similarly.

Since the empty site moves diagonally along the interface of jamming stripe, we propose a method to estimate $p_{S_{1}^{e}}$. We assume that the probabilities of the southwest (northeast) site occupied by the E pedestrian and the N pedestrian are both $1/2$. We assume the probability of the southeast, east-east and south-south site occupied by the E pedestrian are both $p_{E}^{R}$ while the probabilities for the N pedestrian are $p_{N}^{R}$. Similarly, the probabilities of the northwest, west-west and north-north site occupied by the E pedestrian are $p_{E}^{L}$, while the probabilities for the N pedestrian are $p_{N}^{L}$. Thus, the probability of the $S_{1}^{e}$ is

\begin{eqnarray}\label{20}
p_{S_{1}^{e}}&\approx&\frac{1}{2}p_{N}^{R}p_{N}^{R}.
\end{eqnarray}

From FIG. \ref{Figure 9}(b) and FIG. \ref{Figure 10}(b), one can see that $p_{N}^{R}$ approximately equals the probability of $S_{2}^{b}$, i.e., $p_{N}^{R}\approx p_{S_{2}^{b}}$. Similarly, $p_{E}^{R}$, $p_{N}^{L}$ and $p_{E}^{L}$ approximately equal $p_{S_{6}^{b}}$ , $p_{S_{4}^{b}}$ and $p_{S_{8}^{b}}$ respectively. The four probabilities are shown in equation (21)

\begin{eqnarray}\label{21}
p_{E}^{L}&\approx&p_{S_{8}^{b}}=p_{5}+p_{9}+p_{10}+p_{11}+p_{13}+p_{14}+p_{15}+p_{16};
{}\nonumber\\
p_{N}^{L}&\approx&p_{S_{4}^{b}}=p_{1}+p_{2}+p_{3}+p_{4}+p_{6}+p_{7}+p_{8}+p_{12};
{}\nonumber\\
p_{E}^{R}&\approx&p_{S_{6}^{b}}=p_{3}+p_{6}+p_{8}+p_{10}+p_{12}+p_{13}+p_{15}+p_{16};
{}\nonumber\\
p_{N}^{R}&\approx&p_{S_{2}^{b}}=p_{1}+p_{2}+p_{4}+p_{5}+p_{7}+p_{9}+p_{11}+p_{14}.
{}\nonumber\\
\end{eqnarray}

Substituting equations (18)-(21) into (17), we have

\begin{widetext}
\begin{eqnarray}\label{22}
\frac{dp_{1}}{dt}&=&[\frac{1}{2}p_{2}^{r}p_{2}^{r}p_{S_{1}^{b}}q+\frac{1}{2}p_{2}^{r}p_{2}^{r}p_{S_{2}^{b}}\frac{1-q}{2}+\frac{1}{2}p_{2}^{l}p_{2}^{l}p_{S_{4}^{b}}\frac{1-q}{2}]
-[p_{1}q+p_{1}\frac{1-q}{2}+p_{1}\frac{1-q}{2}]
{}\nonumber\\
&=&[\frac{1}{2}(p_{1}+p_{2}+p_{4}+p_{5}+p_{7}+p_{9}+p_{11}+p_{14})^{2}(p_{1}+p_{2}+p_{3}+p_{5}+p_{6}+p_{9}+p_{10}+p_{15})q
{}\nonumber\\& &
+\frac{1}{2}(p_{1}+p_{2}+p_{4}+p_{5}+p_{7}+p_{9}+p_{11}+p_{14})^{3}\frac{1-q}{2}
{}\nonumber\\& &
+\frac{1}{2}(p_{1}+p_{2}+p_{3}+p_{4}+p_{6}+p_{7}+p_{8}+p_{12})^{3}\frac{1-q}{2}]
-[p_{1}q+p_{1}\frac{1-q}{2}+p_{1}\frac{1-q}{2}]=0.
\end{eqnarray}
\end{widetext}

The equations for $p_{2}\sim p_{16}$ can be obtained similarly. However, note that only fifteen of the sixteen equations are independent ones.

The conservation of probability requires that

\begin{eqnarray}\label{23}
\sum_{i=1}^{16}p_{i}=1.
\end{eqnarray}

Now, we have sixteen equations for the sixteen variables. We cannot obtain analytical result of the equations. Instead, we can obtain numerical result.

The average movement of empty site $<v_{mov}>$ in one MCS can be calculated from $p_{1}~p_{16}$

\begin{widetext}
\begin{eqnarray}\label{24}
<v_{mov}>&=&p_{1}(q+2\frac{1-q}{2})+p_{2}(q+3\frac{1-q}{2})+p_{3}(q+\frac{1-q}{2})
+p_{4}\frac{3(1-q)}{2}+p_{5}(2q+\frac{1-q}{2})+p_{6}(q+2\frac{1-q}{2})
{}\nonumber\\& &
+p_{7}\frac{4(1-q)}{2}+p_{8}\frac{2(1-q)}{2}+p_{9}(2q+2\frac{1-q}{2})+2p_{10}q+p_{11}(q+2\frac{1-q}{2})+p_{12}\frac{3(1-q)}{2}
{}\nonumber\\& &
+p_{13}(q+\frac{1-q}{2})+p_{14}(q+3\frac{1-q}{2})+p_{15}(2q+\frac{1-q}{2})+p_{16}(q+2\frac{1-q}{2}).
\end{eqnarray}
\end{widetext}

\begin{figure}
    \begin{center}
    \scalebox{0.3}[0.3]{\includegraphics{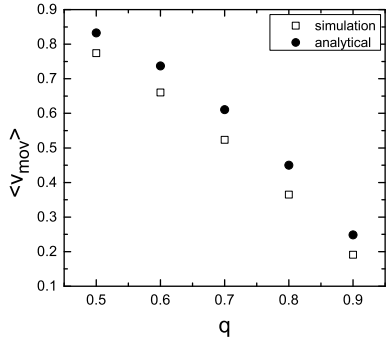}}
    \caption{The average movement of the empty site in one MCS  $<v_{mov}>$ against $q$. The solid and open squares are analytical and simulation  results respectively.}\label{Figure 11}
    \end{center}
\end{figure}

FIG. \ref{Figure 11} compares the simulation results with the analytical results. One can see that the analytical results are in approximate agreement with simulation ones. This is due to that the correlation among the neighbors of empty site. More effective methods are needed for the probability of $S_{1}^{e}$, etc.

\subsection{Open boundary}

\begin{figure}
    \begin{center}
    \scalebox{0.3}[0.3]{\includegraphics{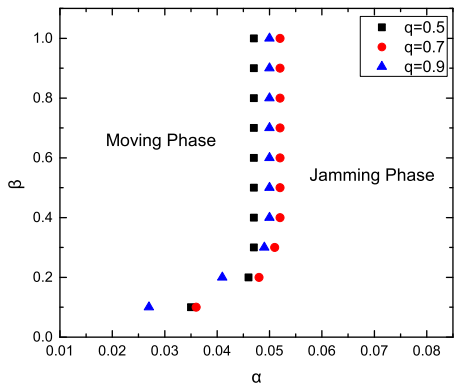}}
    \caption{Phase diagram of the model under open boundary conditions for lattice size $L=100$ .}\label{Figure 12}
    \end{center}
\end{figure}

\begin{figure}
    \begin{center}
    \scalebox{0.3}[0.3]{\includegraphics{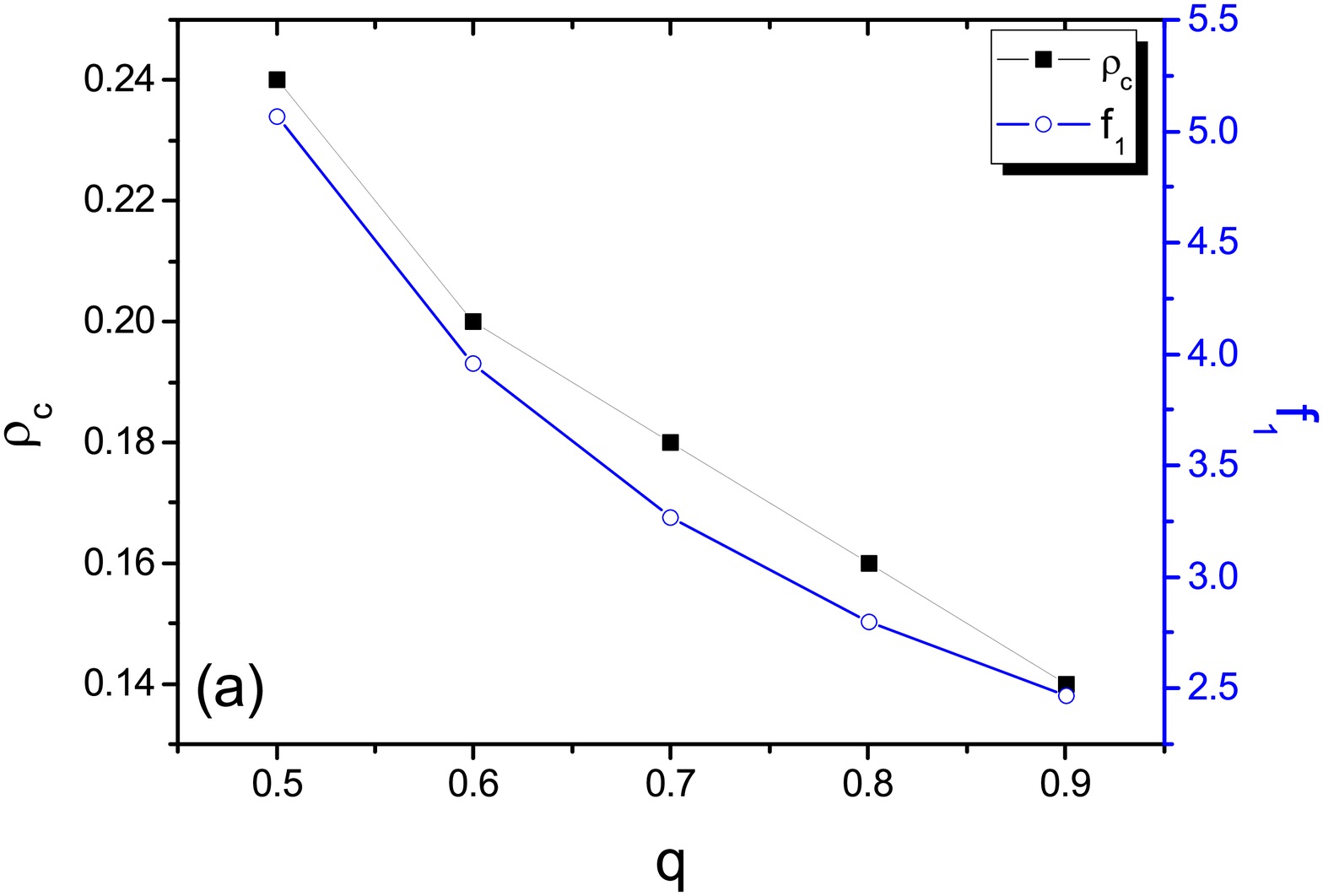}}
    \scalebox{0.3}[0.3]{\includegraphics{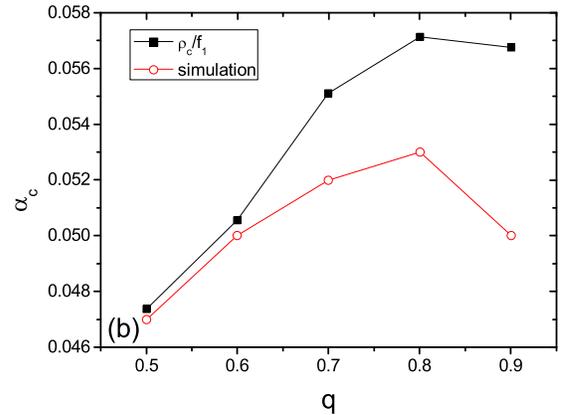}}
    \caption{ (a)The simulation results of $\rho_{c}$ and $f_1$  against $q$ as $\alpha=0.045$, $\beta=1$ and $L=100$. (b) $\alpha_c \approx\rho_{c}/f_1$ and $\alpha_c$ of FIG.\ref{Figure 12} against $q$ .}\label{Figure 13}
    \end{center}
\end{figure}

\begin{figure}
    \begin{center}
    \scalebox{0.3}[0.3]{\includegraphics{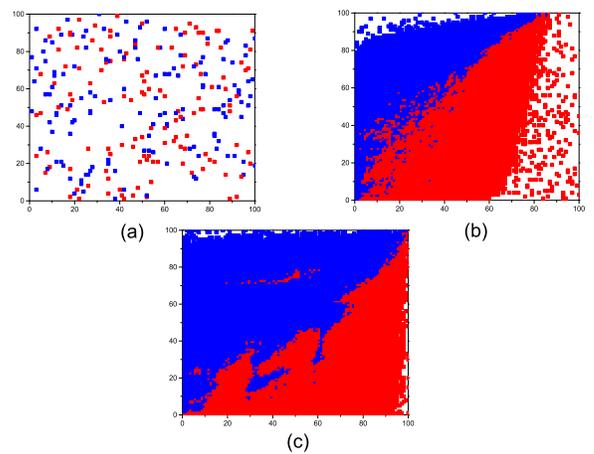}}
    \caption{Three typical configurations of the model with open boundary conditions at $q = 0.7$. The parameters are $L = 100$, $\beta = 1.0$ and (a) $\alpha = 0.01$,(b) $¦Á = 0.056$, (c) $\alpha = 0.2$. The E pedestrian is indicated by blue and the N pedestrian is indicated by red.}
    \label{Figure 14}
    \end{center}
\end{figure}

FIG.\ref{Figure 12} shows the phase diagram under open boundary conditions when $q=0.5,0.7,0.9$. Similar to the result of reference \cite{mod22,mod26}, there are two phases, i.e. the moving phase and the jamming phase.

\begin{figure}
    \begin{center}
    \scalebox{0.3}[0.3]{\includegraphics{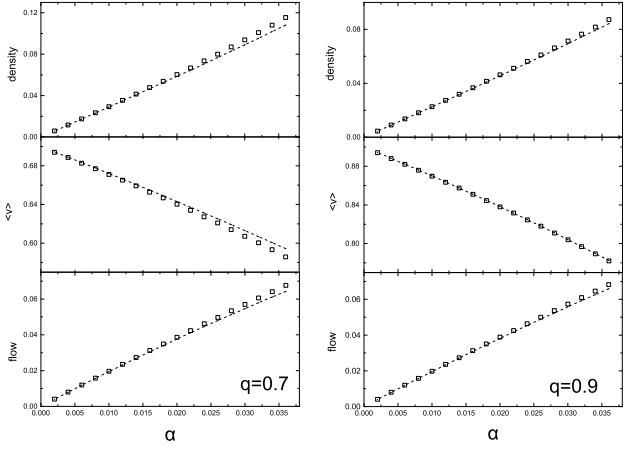}}
    \caption{The global density, the average velocity and the flow in the moving phase at (a) $q=0.7$,(b) $q=0.9$ and $\beta=1$. The simulation results for the global density, average velocity $<v>$ and flow are the average of 100 runs.}
    \label{Figure 15}
    \end{center}
\end{figure}

Three typical configurations for $q=0.7$ are shown in the FIG. \ref{Figure 14}. The typical configuration of the moving phase is similar to that of the periodic boundary condition, where the pedestrians are randomly and uniformly distributed on the lattice (see FIG. \ref{Figure 14}(a)). When approaching the boundary of jamming phase, the jamming state begins to emerge in the system.  The jamming region expands gradually and the system is almost fully occupied by pedestrians (see FIG. \ref{Figure 14}(c)). Different from the result of reference \cite{mod22}, the shape of jamming region becomes diamond. Because the E (N) pedestrian can move out from the up (right) boundary (see FIG. \ref{Figure 14}(b)).
%These features are  similar to that of $q=1$ \cite{mod19,mod22,mod26,mod27,mod28}.

From FIG.\ref{Figure 12}, one can see that the critical injection probability $\alpha_{c}$ increases firstly then decreases with the increase of $q$.
The explanation is as follows. As shown in FIG.\ref{Figure 15}, the density increases almost linearly with $\alpha$, i.e., $\rho \approx \alpha *f_1(q)$.  When $\alpha$ increases to $\alpha_c$, the density exceeds $\rho_{c}$ of period boundary condition and the jam happens, i.e., $\alpha_c \approx \rho_{c}(q)/f_1(q)$.  FIG.\ref{Figure 13}(a) shows the simulation results of $\rho_{c}$ and $f_1$ against $q$. One can see that both $\rho_{c}$ and $f_1$ decrease with the increase of $q$. However the decline rates are different. The decline rate of $f_1$  is greater than that of $\rho_{c}$  when $q<0.8$, while smaller when $q>0.8$. The $\alpha_c \approx \rho_{c}/f_1$ shown in FIG.\ref{Figure 13}(b) agrees with $\alpha_c$ of FIG.\ref{Figure 12} qualitatively.

We study the global density $\rho$, the average velocity $<v>$ and the flow $J$ in the moving phase. Because of the conservation of flow, the flow in the bulk equals the inflow; i.e.

\begin{eqnarray}\label{25}
J&=&\frac{\rho}{2}qp_{f}=\alpha(1-\rho).
\end{eqnarray}

Combining the equations (1)-(3) and (25), we obtain the $\rho$ and $<v>$ for each $\alpha$ and $q$. Then the flow can be calculated by using equation (25). The analytical results are shown in FIG. \ref{Figure 15} and are in good agreement with the simulation results.

\section{Conclusion}

In this paper, the intersecting  pedestrian flow on two-dimensional lattice has been studied. Under periodic boundary condition, the intermediate phase in which some pedestrians could move along the border of jamming stripes were observed. The density where multi stripes begin to appear decreases with the increase of $q$. We have developed a mean field analysis for the moving phase by extending the method of \cite{mod22}. The analytical results agree with the simulation results well. The average velocity of intermediate phase was obtained by the analytical result of a 1D model, which agree with the simulation results well. When the system has only one empty site, the average movement in one MCS was obtained by mean field analysis. There is a little deviation between the simulation and analytical result because of the correlation among the pedestrians. More accurate methods will be developed in the future work.

Under the open boundary conditions, the moving phase and jamming phase were observed. The shape of jamming region changes into diamond. The critical injection probability $\alpha_{c}$ shows nontrivially against $q$. The analytical results for the flow rate, average velocity and density in the moving phase were obtained and agree with the simulation results well.

The analytical methods in the paper could be generalized to the parallel updating models. The model also can be extended to other complex environment such as the subway, classroom.

\subsection{Acknowledgments}

This work is funded by the National Natural Science Foundation of China (Grant Nos. 71671058, 71301042, 71431003), the Doctoral Program of the Ministry of Education (No. 20130111120027), Fundamental Research Funds for the Central Universities£¬ Singapore Ministry of Education Academic Research Fund Tier 2 (Grant No. MOE2013-T2-2-033). The numerical calculations in this paper have been done on the supercomputing system in the Supercomputing Center of University of Science and Technology of China.


\section{References}

\begin{thebibliography}{ref1}
\bibitem{dyn1} D. Helbing, I. Farkas, T. Vicsek, Nature. 407 (2000) 487.
\bibitem{dyn2} M. Schreckenberg, S. D. Sharma, Pedestrian and Evacuation Dynamics, New York, 2002.

\bibitem{con9} D. Helbing, I.J. Farkas, P. Molnar, T. Vicsek, Pedestrian and Evacuation Dynamics, (2002) 21-58.

\bibitem{phen3}S. Seer, N. Brandle, C. Ratti, Transp. Res. Part. C 48 (2012) 212-228.

\bibitem{exp2} W. Daamen, S.P. Hoogendoorn, Transp. Res. Rec, 1828 (2003) 20.
\bibitem{exp3} W.H.K. Lam, J.Y.S. Lee, K.S. Chan, Transp. Res. A 37 (2003) 789.
\bibitem{exp4} D. Helbing,M. Isobe,T. Nagatani, Phys. Rev. E 67 (2003) 067101.
\bibitem{exp5} D. Helbing,L. Buzna, A. Johansson, Transp. Sci. 39 (2005) 1.
\bibitem{exp6} S.P. Hoogendoorn, W. Daamen, Transp. Sci. 39 (2005) 147.
\bibitem{exp7} T. Kretz, A. Gr¨¹nebohm, M. Kaufman, J. Stat. Mech. (2006) P10001.

\bibitem{Helbing1995}D. Helbing, Phys. Rev. E 51 (1995) 3164-3169.

\bibitem{Henderson1} L. F. Henderson, Nature. 229 (1971) 381.

\bibitem{con8} D. Helbing, P. Molnar, Phys. Rev. E 51 (1995) 4282-4286.

\bibitem{con10} A. Nakayama, K. Hasebe, Y. Sugiyama, Phys. Rev. E 71 (2005) 036121.
\bibitem{nakayama}A. Nakayama, K. Hasebe, Y. Sugiyama, Comput. Phys. Commun. 177 (2007) 162-163.
\bibitem{cel11} M. Muramatsu, T. Nagatani, Physica A 275 (2000) 281-291.

\bibitem{cel12} M. Muramatsu, T. Irie, T. Nagatani, Physica A 286 (2000) 377-390.
\bibitem{cel13} K. Takimoto, T. Nagatani, Physica A 320 (2003) 611-621.
\bibitem{cel14} Y. Tajima, T. Nagatani, Physica A 303 (2002) 239-250.
\bibitem{cel15} Y. Tajima, T. Nagatani, Physica A 292 (2001) 545-554.
\bibitem{cel16} Y. Tajima, K. Takimoto, T. Nagatani, Physica A 294 (2001) 257-268.


\bibitem{pengy} Y.C. Peng, C.I. Chou, Comput. Phys. Commun. 182 (2011) 205-208.

\bibitem{exp17} R.Y. Guo, S.C. Wong, H.J. Huang, Physica A 389 (2010) 515-526.
\bibitem{exp18} L.P. Lian, X. Mai, W.G. Song, J. Stat. Mech. (2015) P08024.
\bibitem{mod19} H.J. Hilhorst, C. Appertrolland, J. Stat. Mech. (2012) P06009.

\bibitem{mod27} J. Cividini, C. AppertRolland, H.J. Hilhorst, Europhys. Lett. 102 (2013) 20002.
\bibitem{mod28} J. Cividini , H.J. Hilhorst, C. AppertRolland, J. Phys. A: Math. Theor. 46 (2013) 345002.

\bibitem{mod21} O. Biham, A.A. Middleton, D.A. Levine, Phys. Rev. A 46 (1992) R6124-R6127.
\bibitem{inter phase} R.M. D'Souza, Phys. Rev. E 71 (2005) 066112.
\bibitem{mod22} Z.J. Ding, R. Jiang, B.H. Wang, Phys. Rev. E 83 (2011) 047101.
\bibitem{suiqh} Q.H Sui, Z.J. Ding, R. Jiang, W. Huang, D. Sun, B.H. Wang,
Comput. Phys. Commun. 183 (2012) 547-551.

%\bibitem{cur23} S.P. Hoogendoorn, W. Daamen, Transp. Sci., 39 (2005) 147-159.
\bibitem{mod24} S.A. Janowsky, J.L. Lebowitz, Phys. Rev. A 45 (1992) 618.
\bibitem{mod25} S.A. Janowsky, J.L. Lebowitz, J. Stat. Phys. 77 (1994) 35.
\bibitem{mod26} Z.J. Ding, Z.Y. Gao, J.C. Long, J. Stat. Mech. (2014) P10002.
%\bibitem{mod29} J. Cividini, C. AppertRolland, J. Stat. Mech., (2013) P07015.

\end{thebibliography}
\end{document}